\documentclass[preprint]{revtex4}
\usepackage{epsfig}

\def\beq{\begin{equation}}
\def\eeq{\end{equation}}
\def\beqa{\begin{eqnarray}}
\def\eeqa{\end{eqnarray}}

\def\GeV{\nobreak\,\mbox{GeV}}

\def\nn{\nonumber}

\newcommand{\tnk}{\Theta nK}

\newcommand{\lll}{\lambda_{\Theta}\lambda_{N}\lambda_{K}}
\def\me#1{\langle{#1}\rangle}

\def\bra#1{\langle #1|}
\def\ket#1{| #1\rangle}
\def\qbar{\overline{q}}
\def\gs{g_{\rm s}}
\def\G{{\cal G}}
\def\mixbar{\gs\qbar\sigma\!\cdot\!\G q}

\def\gluoncon{{\displaystyle{\gs^2G^2}}}

\def\xsla{\rlap{/}{x}}

\newcommand{\gev}{\mbox{\rm GeV}}

\begin{document}

\title{\sc Pentaquark Decay Width in QCD Sum Rules}

\preprint{IFIC/05-20}

\author{M. Eidem\"uller\ddag\ \dag\ , F.S. Navarra\ddag\ , M. Nielsen\ddag\ and 
R. Rodrigues da Silva\ddag\ }
\address{\ddag\ Instituto de F\'{\i}sica, Universidade de S\~{a}o Paulo\\
 C.P. 66318,  05315-970 S\~{a}o Paulo, SP, Brazil}
\address{\dag\ Departament de F\'{\i}sica Te\`orica, IFIC,
          Universitat de Val\`encia -- CSIC,\\
Apt. Correus 22085, E-46071 Val\`encia, Spain}

\begin{abstract}
In a diquark-diquark-antiquark picture of the pentaquark we study the decay 
$\Theta \rightarrow K^{+} n$ within the framework of QCD sum rules.
After evaluation of the relevant three-point function, we extract the coupling
$g_{\Theta nK}$ which is directly related to the pentaquark width.
Restricting the decay diagrams to those with color exchange 
between the meson-like and baryon-like clusters reduces the coupling constant 
by a factor of four. 
Whereas a small decay width might be possible for a positive parity pentaquark,
it seems difficult to explain the measured width for a pentaquark with 
negative parity.

\end{abstract} 

\pacs{PACS Numbers~ :~ 12.38.Lg, 12.40.Yx, 12.39.Mk}
\maketitle



\vspace{1cm}
\section{Introduction}

The possible existence of the pentaquark is one of the most exiting 
questions in current particle phenomenology. So far, more than ten experiments
have found evidence for the existence of a pentaquark state
\cite{PosResults}.
However, an almost equal number of experiments do not see any signal at the
mass of the pentaquark \cite{NegResults}. 
The experiments who reported positive results work in a medium energy 
range and have limited statistics. No pentaquark was seen in high energy
experiments, however, the production mechanism is unclear and it is very 
difficult give an estimate for the production rate. Higher statistics
experiments will be needed to clarify the existence of this new hadronic state.
A review on the present experimental status can e.g. be found in
\cite{dzi}.

These investigations have triggered an intense theoretical activity and for
an overview on the theoretical status we refer to reader to
\cite{theory}.
One of the most puzzling characteristics of the pentaquark is its extremely
small width (much) below 10 MeV which poses a serious challenge to all theoretical
models.  Indeed, in the conventional  uncorrelated quark model the expected 
width is of the order of several hundred MeV, since the strong decay 
$\Theta^{+} \, \rightarrow K N$ is Okubo-Zweig-Iizuka (OZI) super-allowed. 
Many explanations for this narrow width have been advanced 
\cite{carlson,Stech,oga,ricanari}.  
A suggestive way to explain the small width is by the assumption of
diquark clustering \cite{Jaffe:2003sg}.
The formation of diquarks presents an important concept
and has direct phenomenological impact \cite{Diquarks}.

In this work we calculate the pentaquark decay width within the framework of QCD
sum rules. The basis of the sum rules was laid
in \cite{SRbasis} and their extension to baryons was developed in
\cite{SRbaryons}.
The assumptions of the model are incorporated by an appropriate current. 
The sum rules are directly based on QCD and keep the analytic dependence
on the input parameters.
Several sum rule calculations have been performed for the mass
of the pentaquark containing a strange \cite{zhu,mnnr,oka,eiden} or charm 
\cite{suhong_charm} quark. These calculations are based on two-point functions
with different interpolating currents. Surprisingly, all these determinations
give similar masses with reasonable values. 
A common problem of all determinations
is the large continuum contribution which has its origin in the high dimension
of the interpolating currents and results in a large dependence on the
continuum threshold. Another problem is the irregular behavior of the operator 
product expansion (OPE), which is  dominated by higher dimension operators and 
not by the perturbative term as it should be.

Here we present for the first time a sum rule determination for decay width
based on a three-point function for the decay $\Theta\to n K^+$. In this way
we can extract the coupling $g_{\Theta n K}$ which is directly related
to the pentaquark width. To describe the pentaquark we use the
diquark-diquark-antiquark model with one scalar and one pseudoscalar
diquark in a relative S-wave.

In \cite{oga,ricanari} it has been argued that such a small
decay width can only be explained if the pentaquark is a genuine 5-quark
state, i.e., it contains no color singlet meson-baryon contributions and thus 
color exchange is necessary for the decay.
They start from simple observations concerning symmetries of the 
quark currents and the properties of the basic decay diagram as shown in Fig. 1.
The analysis presented both in \cite{oga} and in \cite{ricanari} is only 
qualitative  
and arrives at the conclusion that, for a particular choice of the pentaquark 
current and assuming that it is a genuine 5-quark state, the decay width is 
given by
$\Gamma_{\Theta} \simeq  \alpha^2_s \, <0|\overline{q} q|0>^2$.
In this expression $\alpha^2_s$ appears due to a hard gluon exchange and is 
therefore small. The narrowness of the pentaquark width can then be attributed to 
chiral symmetry breaking and the non-trivial color structure of 
the pentaquark which requires the exchange of, at least, one gluon.
In this work we will also test quantitatively the hypothesis put forward in
\cite{oga,ricanari} to see whether this mechanism is sufficient 
to explain the small width. The effect of chiral symmetry breaking
for two chirally different diquarks in a relative S-wave was estimated
in \cite{Stech}. It was concluded that chiral breaking alone is not sufficient
to explain a very small pentaquark width and our investigation will support
this finding.


\section{The correlation function}

Whereas the QCD sum rule determinations for the mass of the pentaquark are based
on two-point functions \cite{zhu,mnnr,oka,eiden}, an investigation of the
decay width requires a three-point function which we define as
\begin{eqnarray}
\label{correl}
\Gamma(p,p^{,}) &=&
\int d^4x \, d^4y  \, e^{-iqy} \, e^{ip^{,}x} \,\Gamma(x,y)\,, \nn\\
\Gamma(x,y) &=& \langle 
0|T\{\eta _N(x)j_{K}(y) \bar{\eta}_{\Theta}(0)\}|0 \rangle\,,
\end{eqnarray}
where $\eta_N$, $j_{K}$ and $\eta_{\Theta}$ are the interpolating fields associated 
with neutron, kaon and  $\Theta$, respectively. 

\subsection{The phenomenological side}

We now  consider the expression (\ref{correl}) in terms of hadronic degrees of freedom 
and write the phenomenological  side of the sum rule. 
Though in \cite{oka} it has been argued that a scalar-pseudoscalar diquark 
model is more likely related to a negative parity pentaquark, in this work
we make no assumption about the parity of the $\Theta$ and investigate
the consequences for both cases.
Treating the kaon as a pseudoscalar particle, the interaction between the three 
hadrons is described by the following Lagrangian density:
\begin{eqnarray}
\label{lag}
{\cal L}&=&ig_{\Theta nK}\bar{\Theta}\gamma_5 Kn\quad\mbox{for $P=+$}\nn\\
{\cal L}&=&ig_{\Theta nK}\bar{\Theta}Kn\quad \ \ \ \mbox{for $P=-$}
\end{eqnarray} 
Writing the correlation function (\ref{correl}) in momentum space and inserting 
complete sets of hadronic states we obtain
\begin{equation}
\label{vfen}
\Gamma(p,p')= \sum_{s,s'} \, - i \, 
\frac{
\langle 0|\eta _N|n(p',s')\rangle  V(p,p') \langle K(q)|j_{K}|0\rangle
\langle \Theta(p,s)
|\bar{\eta}_{\Theta}|0 \rangle}
{(p'^{2}- m_{N}^2)(q^2- m_{K^{+}}^2) (p^2- m_{\Theta}^2)},
\end{equation}
with
\begin{eqnarray}
\label{MatrixElements}
- i \, V(p,p') &=& < n(p',s') | \Theta(p,s) K(q) >\,, \nn\\
\langle 0|\eta _N|n(p',s')\rangle &=& \lambda_{N} u^{s'}(p')\,,\nn\\
\langle K(q)|j_{K}|0\rangle &=& \lambda_{K}\,,\nn\\
\langle \Theta(p,s)|\bar{\eta}_{\Theta}|0 \rangle &=&
\lambda_{\Theta}\bar{u}^{s}(p)\quad \ \ \ \ \ \mbox{for $P=+$}\nn\\
\langle \Theta(p,s)|\bar{\eta}_{\Theta}|0 \rangle &=&
-\lambda_{\Theta}\bar{u}^{s}(p)\gamma_{5}\quad \mbox{for $P=-$}
\end{eqnarray}
where the spinors  are normalized according to
\begin{equation}
\label{espinor}
\sum_{s=1,2} \, u^{s}(p)\bar{u}^{s}(p) =\not{\!p}+m.
\end{equation}
Using the simple Feynman rules derived from (\ref{lag}) we can rewrite 
$ V(p,p')$ as
\begin{eqnarray}
V(p,p') &=& - \,  g_{\Theta nK} \bar{u}^{s'}(p') \gamma_5 u^{s}(p)
\quad\mbox{for $P=+$}\nn\\
V(p,p') &=& - \, g_{\Theta nK} \bar{u}^{s'}(p') u^{s}(p)\quad\ \ \ \mbox{for $P=-$}
\end{eqnarray}
The coupling constants $\lambda_{N}$ and  $\lambda_{\Theta}$ can  be
determined from the QCD sum rules of the corresponding two-point functions.
$\lambda_{K}$ is related to the kaon decay constant through 
\begin{equation}
\label{decaek}
\lambda_{K}= \frac{f_{K}m_{K}^2}{m_u+m_s}\,.
\end{equation}
Combining the expressions above we arrive at
\begin{equation}
\label{vfen2}
\Gamma(p,p')= -ig_{\Theta nK}
\lambda_{\Theta}\lambda_{N}\lambda_{K}
\frac{
(\not{\!p'}+m_{N})
(\not{\!p} \pm m_{\Theta})\gamma_{5}}
{(p'^2- m_{N}^2)(q^2 - m_{K^+}^2)
(p^2- m_{\Theta}^2)} \,\, + \,\, \mbox{continuum}
\end{equation}
where the upper (lower) sign corresponds to a negative (positive) parity pentaquark.   
Finally, writing all the Dirac structures explicitly, the phenomenological 
correlator is given by
\begin{equation}
\label{vfen3}
\Gamma_{phen}(p,p')=
\frac{
-g_{\Theta nK}
\lambda_{\Theta}\lambda_{N}\lambda_{K}}
{(p'^2- m_{N}^2)(q^2- m_{K^{+}}^2)
(p^2- m_{\Theta}^2)}
\Gamma _E  \,\, + \,\, \mbox{continuum}
\end{equation}
with
\begin{equation}
\Gamma _E=
\sigma^{\mu\nu}\gamma _{5}q_{\mu}p'_{\nu}
-im_{N}\not{\!q}\gamma _5
+i( \pm m_{\Theta}+m_{N})\not{\!p'}\gamma _5
+i(p'^2 \pm m_{\Theta}m_{N}-qp')\,.
\label{estrut}
\end{equation}
We shall work with the 
$\sigma^{\mu\nu}\gamma _{5}q_{\mu}p'_{\nu}$ structure  because, as it was shown 
in \cite{gamma5}, 
this structure gives results which are less sensitive to the coupling scheme 
on the phenomenological side, i.e., to the choice of a pseudoscalar or 
pseudovector coupling between the kaon and the baryons.
The continuum part contains the contributions of all possible excited states.  
They require a special treatment and we will discuss these terms 
separately after presenting the theoretical contributions.


\subsection{The theoretical side}

We now come back to (\ref{correl}) and write the interpolating fields  in terms of 
quark degrees of freedom as
\begin{eqnarray}
\label{currents}
j_{K}(y)&=&\bar{s}(y)i\gamma_{5}u(y)\,,\nn\\
\eta _{N}(x)&=&\epsilon^{abc}
({d}^{T}_a(x)C\gamma _{\mu}d_{b}(x))
\gamma _5 \gamma ^{\mu}u_c(x)\,,\nn\\
\bar{\eta}_{\Theta}(0)&=&
-\epsilon^{abc}\epsilon^{def}\epsilon^{cfg}
{s}^{T}_{g}(0)C
[\bar{d}_e(0)\gamma _5C\bar{u}^{T}_d(0)]
[\bar{d}_b(0)C\bar{u}^{T}_a(0)]\,.
\end{eqnarray}
The current of the pentaquark incorporates the assumption of a
diquark-diquark-antiquark system with a scalar and pseudoscalar diquark.
Inserting these currents into  (\ref{correl}), the resulting expression can 
be written in the following form:
\begin{eqnarray}
\label{gamma1}
\Gamma_{th}(x,y)&=&
2i\epsilon^{abc}\epsilon^{def}\epsilon^{cfg}
\epsilon^{a'b'c'} \widetilde{\Gamma}(x,y)\,,\nn\\
\widetilde{\Gamma}(x,y)&=& [N_2(x) - N_1(x)]K(y)\,,
\end{eqnarray}
with
\begin{eqnarray}
\label{n1}
N_1(x)&=& \gamma _5\gamma ^{\beta}
S_{c'd}(x)C S^{T}_{a'e}(x)C \gamma _{\beta} S_{b'b}(x)\gamma _5\,,\nn\\
N_2(x) &=& \gamma _5\gamma ^{\beta}
S_{c'd}(x)\gamma _5 C S^{T}_{a'e}(x)C \gamma _{\beta} S_{b'b}(x)\,,\nn\\
K(y)&=& C S^{T}_{ha}(y)C \gamma _{5}C S^{T}_{gh}(-y,m_s) C\,.
\end{eqnarray}
In order to proceed with the  evaluation of the correlator
(\ref{correl}) at the quark
level, we  need  the quark propagator in the
presence of quark and gluon condensates. Keeping track of the terms
linear in the quark mass and taking into account quark and gluon 
condensates, we have 
\beqa
S_{ab}(x)&=&\bra{0} T[q_a(x)\overline{q}_b(0)]\ket{0}={i\delta_{ab}\over2
\pi^2x^4}\xsla-{m_q\delta_{ab}\over4\pi^2x^2}-{i\over32\pi^2x^2}t^A_{ab}
\gs G^A_{\mu\nu}
(\xsla\sigma^{\mu\nu}+\sigma^{\mu\nu}\xsla)
\nonumber\\
&-&{\delta_{ab}\over12}\me{\qbar q}
-{m_q\over32\pi^2}t^A_{ab}\gs G^A_{\mu\nu}\sigma^{\mu\nu}\ln(-x^2)
+{i\delta_{ab}\over48}m_q\me{\qbar q}\xsla-{x^2\delta_{ab}
\over2^6\times3}\me{\mixbar}
\nonumber\\
&+&{ix^2\delta_{ab}\over2^7\times3^2}m_q\me{\mixbar}\xsla
-{x^4\delta_{ab}\over2^{10}\times3^3}\me{\qbar q}\me{\gluoncon}\, 
-{x^2 ln(-x^2) m_q \delta_{ab}\over2^{9}\times3 \pi^2}\me{\qbar q}\me{\gluoncon} 
\label{prop}
\eeqa
where we have used the factorization approximation for the multi-quark
condensates and we have used the fixed-point gauge. Inserting (\ref{prop}) into 
Eqs. (\ref{n1}) and  these into  (\ref{gamma1}),
we arrive at a complicated function which is given schematically 
by the sum of the diagrams of Fig. 2. For any given  Dirac structure 
this function can be written in a factorized form:
\begin{equation}
\Gamma_{th}(x,y)= H(x^2) \cdot L(y^2)\,.
\label{gammahl}
\end{equation}
After Fourier transformation, this leads to a similar separation
in momentum space. The diagrams of Fig. 2 can then
be expressed in terms of ${\widetilde H}(p'^2)$ and
${\widetilde L}(q^2)$.


\subsection{The continuum part and pole-continuum transitions}

Let us consider the phenomenological side (\ref{vfen3}) and,  
following \cite{ioffe3}, rewrite it  generically as a double dispersion 
relation:
\begin{equation}
\Gamma(q^2, p^2, p'^2) \, = \, \int_{0}^{\infty} \,  d s
\int_{0}^{\infty} \, d u \, 
\frac{\rho(s,u,p^2)}{(s -  p'^2) (u - q^2)}\,.
\label{double_ioffe}
\end{equation}
The double discontinuity can be written as the sum:
\begin{eqnarray}
\rho(s,u,p^2) \, & = & \, a(p^2) \delta(s - m^2_N) \delta(u - m^2_K) \, + \, 
b_1(u,p^2) \delta(s - m^2_N) \, \theta(u - m^2_{K^{*}})  \,  \nonumber\\
&+& \, b_2(s,p^2) \delta(u - m^2_K) \, \theta(s - m^2_{N^{*}}) \, + \, 
\rho_{cc} (s,u,p^2) \theta(s - s_0) \theta(u - u_0)\,,
\label{discont_ioffe}
\end{eqnarray}
where the continuum thresholds are defined as
\begin{equation}
s_0  \, = \, ( m_N \, + \, \Delta_N )^2  \,\,\,\,\,\,   \mbox{GeV}^2 
\,\,\,\,\,\,\,\,\,\,
u_0  \, = \, ( m_K \, + \, \Delta_K )^2   \,\,\,\,\, \mbox{GeV}^2\,.
\label{threshold}
\end{equation}
The terms proportional to $b_1$ and $b_2$ represent pole-continuum transitions
where the pentaquark (in the ground state or in an excited state) decays into one 
ground state nucleon and one exited kaon or vice versa.  
Inserting (\ref{discont_ioffe}) into (\ref{double_ioffe}) we can write
\begin{equation}
\Gamma(q^2, p^2, p'^2) \, = \, \Gamma_{pp}(q^2, p^2, p'^2) \, + \, 
\Gamma_{pc1}(q^2, p^2, p'^2) \, + \, \Gamma_{pc2}(q^2, p^2, p'^2) \, + \,  
\Gamma_{cc}(q^2, p^2, p'^2)\,,
\label{soma_ioffe}
\end{equation}
where  $\Gamma_{pp}$ comes from the first term in (\ref{discont_ioffe}) and 
stands for the pole-pole part. The coefficient $a(p^2)$ is obtained from  
Eq. (\ref{vfen3}) and we get
\begin{equation}
 \Gamma_{pp}(q^2, p^2, p'^2)  =
- \frac{g_{\tnk}\lll}{(p^2 - m_{\Theta}^2) (p'^2 -  m^2_N) (q^2 - m^2_K)}\,.
\end{equation} 
The continuum-continuum term $\Gamma_{cc}$ can be obtained as usual, 
with the assumption of quark-hadron duality, from the double dispersion integral
(\ref{double_ioffe}) using the theoretical expressions
$\rho_{cc}(s,u,p^2) \, = \, \rho_{th}(s,u,p^2)$.
We can also write a double dispersion integral for   
$\Gamma_{th}(q^2,p^2,p'^2)$ and, because of duality, 
$\Gamma_{cc}$ may be  transferred to the theoretical side. This  only changes 
the integration  limits, so that the final theoretical side or right-hand side of the 
sum rule reads:
\begin{equation}
\Gamma_{rhs}(q^2, p^2, p'^2) \, = \, \Gamma_{th}(q^2, p^2, p'^2) \, - \, 
\Gamma_{cc}(q^2, p^2, p'^2) \, = \, 
\int_0^{s_0} \,  d s \, \int_0^{u_0} \,  du   \, \frac{\rho_{th}(s,u,p^2)  } 
{(s -  p'^2) (u - q^2)}\,.
\label{gammathf}
\end{equation}
The pole-continuum transition terms are contained in $\Gamma_{pc1}$ and
$\Gamma_{pc2}$. They can be explicitly written as
\begin{eqnarray}
\label{gammapc1}
\Gamma_{pc1}(q^2, p^2, p'^2) &=& \int_0^{\infty} \, d s \, \int_{m^2_{K^{*}}}^{\infty}
 \, \frac{b_1(u,p^2) 
\, \delta(s - m^2_N) \,  du}
{(s -  p'^2) (u - q^2)} =  \int_{m^2_{K^{*}}}^{\infty}  
\, \frac{b_1(u,p^2)  \,  du}
{(m^2_N  - p'^2) (u - q^2)}\,,\nn\\
\Gamma_{pc2}(q^2, p^2, p'^2) &=& \int_{m^2_{N^{*}}}^{\infty} \, d s \,   
\int_0^{\infty} \, \frac{b_2(s,p^2) 
 \, \delta(u - m^2_K) \,  du}
{(s -  p'^2) (u - q^2)} =  \int_{m^2_{N^{*}}}^{\infty} \,
 \frac{b_2(s,p^2) \,  d s}
{(m^2_K - q^2) (s -  p'^2)}\,.
\end{eqnarray}
In a usual three-point sum rule the quark lines in Fig. 1 connect all three
particles, thus forming a triangle graph. After a double Borel transformation
the continuum parts are exponentially suppressed compared to the pole 
contribution. One can then safely make the assumption of quark-hadron duality
and parametrise the continuum by the double discontinuity of the theoretical 
part. However, as was first noticed in \cite{ioffe3}, diagrams where only two particles
are connected contain a contribution in the pole-continuum transitions which is
not exponentially suppressed compared to the pole contribution, even after
double Borel transform. Therefore these contributions can be as large 
as the pole part and must be explicitly included in the sum rules.
Since there is no theoretical tool to calculate the unknown
functions $b_1(u,p^2)$ and $b_2(s,p^2)$ explicitly, one has to employ a
parametrisation for these terms. We will use two different parametrisations:
one with a continuous function for the $\Theta$ and one where the pole term
is singled out. The difference between the two parametrisations should give
an indication about the systematic error.
In the analysis it will turn out that the pole-continuum terms are indeed
of the same order as the pole contribution and should not be neglected.

\smallskip 

{\bf Parametrisation A:}

We shall assume that the functions $b_1$ and $b_2$ have the following form: 
\begin{eqnarray}
b_1(u,p^2) &=& \widetilde{b_1}(u) \int^{\infty}_{m_{\Theta}^2} d \omega
\frac{b_1(\omega)}{\omega - p^2}  \,,  \nonumber \\
b_2(s,p^2) &=& \widetilde{b_2}(s) \int^{\infty}_{m_{\Theta}^2} d \omega
\frac{b_2(\omega)}{\omega - p^2} \,,
\label{b1b2}
\end{eqnarray}
with continuous functions $b_{1,2}(w)$, starting from $m^2_\Theta$.
The functions $\widetilde{b_1}(u)$ and $\widetilde{b_2}(s)$ describe the excitation 
spectra of the kaon and the nucleon, respectively. From experimental data we know that 
the nucleon has a very well established first excitation, the Roper resonance  
($m_{N^{*}} = 1440$ MeV), which is relatively far from the ground state. This suggests that 
the nucleon pole-continuum transitions will be saturated by the nucleon-Roper  
transitions. In order 
to simplify the calculations, we shall take advantage of this fact and use 
$\widetilde{b_2}(s) = \delta(s - m_{N^{*}}^2)$.
In the case of the kaon, surprisingly, no pseudoscalar higher excitation has been 
observed so far. Since the kaon excitations seem not to prefer any particular mass, 
we can not use the simplification applied above to the nucleon and therefore no 
assumption will be made on $\widetilde{b_1}(u)$. After Borel transform,
the pole-continuum term contains one unknown constant factor which can be determined
from the sum rules.

\smallskip

{\bf Parametrisation B:}

In order to investigate the role played by the 
$\Theta$ continuum, we shall now explicitly force the 
phenomenological side to contain only the pole part of the $\Theta$, both in the pole-pole 
term and in the pole-continuum terms. 
This can formally  be done by choosing
$b_1(\omega) = b_2(\omega) = \delta(\omega - m_{\Theta}^2)$.
The functions then read:
\begin{eqnarray}
b_1(u,p^2) &=& \frac{\widetilde{b_1}(u)}{m_\Theta^2-p^2}\,,\nn\\
b_2(s,p^2) &=& \frac{\widetilde{b_2}(s)}{m_\Theta^2-p^2}\,.
\end{eqnarray}
In this case we have the $\Theta$ in the ground state and leave an open nucleon
spectrum.
Again, in the final expressions this gives an additional constant which can
be calculated.


\section{The sum rule}

The sum rule  may be written inserting  (\ref{gamma1}) into  (\ref{correl}) and  
identifying it with (\ref{vfen3}).  As can be seen explicitly in (\ref{estrut}), each  
side of this identity contains a sum of different Dirac structures. The sum rule implies 
that the coefficients of each Dirac structure are equal both in the phenomenological side and
in the theoretical side and therefore a sum rule is actually a set of equations.  
In principle we could work with any of the Dirac structures. As mentioned above, 
we shall work with the $\sigma^{\mu\nu}\gamma _{5}q_{\mu}p'_{\nu}$ structure.

In order to suppress the condensates of higher dimension and at the same time
reduce the influence of higher resonances  we may perform on both sides of the sum rule  
a  standard Borel  transform \cite{SRbasis}: 
\beq
\Pi (M^2) \equiv \lim_{n,Q^2 \rightarrow \infty} \frac{1}{n!} (Q^2)^{n+1} 
\left( - \frac{d}{d Q^2} \right)^n \Pi (Q^2)\,,
\label{borel}
\eeq
($Q^2 = - q^2$) with the squared Borel mass scale $M^2 = Q^2/n$ kept 
fixed in the limit.

The above formulas are quite general. In order to proceed with the numerical analysis we
have to decide  how many Borel transforms we shall perform. In the case of the two-point 
function there is only one four momentum and thus only a single Borel transform is possible. 
In  the case of the three-point function considered here, there are two independent momenta 
and we may perform either a single or a double Borel transform.  
If we were interested in computing the vertex form factor we would necessarily
need  to know the momentum dependence of the vertex function and this would imply making a  
Borel transform in two momentum variables (for example in $p$ and in $p'$), leaving the 
other momentum ($q$) free. 
Since here we are mostly interested in the coupling constant we have more options
which, in principle, should lead to the same result.  We shall 
follow the procedure adopted in the past in similar situations. 
Following \cite{SRbasis} and \cite{thiago} we first consider the choice:
\begin{equation}
\mbox{(I)} \,\,\,\,\,\,\,\,\,\,   q^2 \, = \, 0  \,\,\,\,\,\,\,\,\   p^2 \, = \, p'^2   
\end{equation}
and perform a single Borel transform: $p^2 \, = \, - P^2$ and $ P^2 \rightarrow M^2$. In this
case we take $m_K^2  \simeq 0$ and single out the $1/q^2$-terms. We may call
this scheme the ''Kaon-Pole''-method.
The second choice is:
\begin{equation}
\mbox{(II)} \,\,\,\,\,\,\,\,\,\, q^2 \, \neq  \, 0  \,\,\,\,\,\,\,\,\   p^2 \, = \, p'^2\,. 
\end{equation}
Here we perform two Borel transforms: $p^2 \, = \, - P^2$ and $ P^2 \rightarrow M^2$ 
and also $q^2 \, = \, - Q^2$ and $ Q^2 \rightarrow M^{'2}$. 
We have also considered the choice $ q^2 \, = \, p^2 \, = \, p'^2 \, = - P^2 $, 
performing one single Borel transform ($ P^2 \rightarrow M^2$). This  
procedure was first advanced in \cite{nari86},  used later sometimes 
and has the advantage of simplifying the calculations. However, in the present calculation 
we were not able to find  a stable sum rule. 
Moreover the equal momenta choice is less justified than the others, 
because when we set two momenta squared equal, this bears some connection with the masses or 
virtualities of the particles in the vertex. From this point of view, setting $p^2=p^{'2}$ is 
quite reasonable, since the masses of the on-shell  $\Theta$ and nucleon  are not so different. 
On the other hand, the kaon mass squared is much smaller (and might even be set to zero) 
than $p^2$ and thus $q^2$ should not be close to it.
Therefore in what follows the equal momenta 
choice will not be further considered.

In both  cases considered above,  there 
is no need of making extrapolations in order to obtain the coupling. 
Introducing the notation  
$G = - g_{\Theta nK}\lambda_{\Theta}\lambda_{N}\lambda_{K}$  
and  using (I) and (II) we obtain the following sum rules:

\smallskip

{\bf Method I: Kaon-Pole}

\begin{eqnarray}
\label{soma1}
\Gamma_{pp}(M^2)+\Gamma_{pc2}(M^2)&=&
\int_0^{s_0}  ds  \,\, \rho_{th}(s) \,\,
e^{-s/M^{2}}\,,\quad\mbox{with}\nn\\
\Gamma_{pp}(M^2) &=& G \frac{
e^{-m^2_{\Theta}/M^{2}}  
-e^{-m^2_{N}/M^{2}}  
}
{m^2_{\Theta} - m^2_N}\,,
\end{eqnarray}
and  for the pole-continuum part we obtain
\begin{eqnarray}
\Gamma_{pc2}(M^2) &=& A \,  e^{-m^2_{N^{*}}/M^{2}}
\quad \mbox{for parametrisation A}\nn\\
\Gamma_{pc2}(M^2) &=& A \,  e^{-m^2_\Theta/M^{2}}
\quad \ \mbox{for parametrisation B}
\end{eqnarray}
In both parametrisations the term $\Gamma_{pc1}$ is exponentially suppressed
and, as discussed in \cite{ioffe3}, has been neglected. $A$ is an unknown 
constant and can be determined from the sum rules.

\smallskip

{\bf Method II: Double Borel}

\begin{eqnarray}
\label{soma2}
\Gamma_{pp}(M^2,M^{'2})+\Gamma_{pc2}(M^2,M^{'2}) &=& 
 \int_0^{u_0} \, du \, \int_0^{s_0} \, ds \,  \rho_{th}(s,u) 
 \, e^{-s/M^{2}}
 \, e^{-u/M^{'2}}\,,\quad\mbox{with}  \nn\\
\Gamma_{pp}(M^2,M^{'2}) &=& G \,  e^{-m^2_K/M^{'2}} \,
\frac{e^{-m^2_{\Theta}/M^{2}}  
-e^{-m^2_{N}/M^{2}}}  
{m^2_{\Theta} - m^2_N}
\end{eqnarray}
and
\begin{eqnarray}
\Gamma_{pc2}(M^2,M^{'2}) &=& A \, e^{-m^2_K/M^{'2}} \, e^{-m^{2}_{N^*}/M^{2}}
\quad \mbox{for parametrisation A}\nn\\
\Gamma_{pc2}(M^2,M^{'2}) &=& A \, e^{-m^2_K/M^{'2}} \, e^{-m^{2}_{\Theta}/M^{2}}
\quad \ \mbox{for parametrisation B}
\end{eqnarray}
Also in this case $\Gamma_{pc1}$ is exponentially suppressed.


\section{Results}

\subsection{Numerical input}

In this section we determine the coupling constant $g_{\Theta nK}$.
It is directly related to the experimentally measured $\Theta$ decay width through
\begin{equation}
\label{razao}
\Gamma_{\Theta}= \frac{1}{8\pi m_{\Theta}^3}g_{\Theta nK}^2
[(m_N \mp m_{\Theta})^2-m_K^2] \sqrt{\lambda(m_{\Theta}^2,m_{N}^2,m_K^2)}\,,
\end{equation}
where the upper (lower) sign is for positive (negative) parity and
\begin{equation}
\lambda(m_{\Theta}^2,m_{N}^2,m_K^2)=
(m_{\Theta}^2+ m_{N}^2 -m_K^2)^2-4m_{\Theta}^2m_{N}^2\,.
\end{equation}
The hadronic masses are
$m_N = 938$ MeV, $m_{N^*}=1440$ MeV,  $m_K = 493$ MeV and $m_{\Theta} = 1540$ MeV. 
A decay width of $\Gamma_{\Theta}= 10$ MeV then corresponds to the following
coupling constants:
\begin{eqnarray}
\label{gexp}
g_{\Theta nK} &\simeq& 3.0 \quad \ \mbox{for $P=+$}\nn\\
g_{\Theta nK} &\simeq& 0.43 \quad \mbox{for $P=-$}
\end{eqnarray}
For each of the sum rules above (Eqs. (\ref{soma1}) and (\ref{soma2}))   
we can take the derivative with respect to $1/M^2$ and in this way 
obtain a second sum rule. In each case 
we have thus a system of two equations and two unknowns ($G$ and $A$) which 
can then be easily solved. The results will 
depend on the numerical choices for all input parameters, including
the strange  quark mass, the condensates, the hadron masses and 
the choices of the continuum thresholds for the 
nucleon, $s_0$, and for the kaon, $u_0$. 

In the numerical analysis of the sum rules we use the following values for the 
condensates:  $\me{\qbar q}=\,-(0.23\pm 0.02)^3\,\GeV^3$,
$\langle\overline{s}s\rangle\,=0.8 \, \me{\qbar q}$, $ <\bar{s} g_s {\bf\sigma.G}s > 
= m_0^2 \, \me{\bar{s}s}$ with $m_0^2=0.8\,\GeV^2$ and $\me{\gluoncon}=0.5~\GeV^4$.
The gluon condensate has a large error of about a factor 2, but its influence
on the analysis is relatively small. 
The couplings constants $\lambda _{N}$ and $\lambda _{\Theta}$ are taken 
from the corresponding two-point functions:
\begin{equation}
\lambda _{N}= (2.4 \pm 0.2)  \times 10^{-2} \, \mbox{GeV}^3\,, 
\,\,\,\,\,\,\,\,\,\,
\lambda _{\Theta}= (2.4 \pm 0.3) \times 10^{-5} \, \mbox{GeV}^6\,.
\label{lambdas}
\end{equation}
The coupling $\lambda _{K}$ is obtained from (\ref{decaek}) with $f_K = 160$ MeV,
$m_s = 100$ MeV and $m_u = 5$ MeV: 
\begin{equation}
\lambda _{K}=0.37 \, \mbox{GeV}^2 \,.
\label{lambdak}
\end{equation}

\subsection{Color-connected and color-disconnected diagrams}

As can be seen in Fig. 1, the generic decay diagram in terms of 
quarks has two  ''petals'', one associated with  the kaon and the other with the 
nucleon. In Fig. 2 this picture is completed with the inclusion of  
all the relevant 
condensates. Among these OPE diagrams there are two distinct subsets. 
In the first 
(from 2a to 2g) there is no gluon line connecting the petals and therefore 
no color 
exchange. A diagram of this type we call color-disconnected. In the second 
subset 
of diagrams (2h, 2i and 2j) we have color exchange. If there is no color 
exchange, the 
final state containing two color singlets was already present in the 
initial state, 
before the decay, as noticed in \cite{morimatsu}. 
In this case the pentaquark had a component similar to a  $K-n$ molecule. 
In the second case the pentaquark was a genuine 5-quark  state with a
non-trivial color structure.
We may call this type of diagram a color-connected (CC) one. In our analysis we 
write sum rules for both cases: all diagrams and only color connected. The former 
case is standard in QCDSR calculations and therefore we shall omit details and 
present only the results. The latter case implies that 
 the pentaquark is a genuine
5-quark state and the evaluation of $g_{\Theta nK}$ will thus be based
only on the CC diagrams. In this way we can test whether the assumption of the 
pentaquark being a genuine 5-quark state is sufficient to explain such a 
small decay width.

\subsection{Numerical analysis}

In order to proceed with the numerical analysis  
we must choose a sum rule window
for the Borel parameters $M$ and $M'$. To ensure the convergence of
the OPE we use values above 1 GeV. The upper limit is given by the condition
that the continuum contribution should not be much larger than 50\%.
Thus we use a range of $1\,\gev^2 < M^2,M'^2 < 1.5 \,\gev^2$.
Since the strange mass is small, the dominating diagram is Fig. 2b
of dimension three with one quark condensate. 

We have found out that the contribution from the pole-continuum part is
of a similar size as the pole part. For lower values of $M$ around
1 GeV$^2$, the pole contribution dominates, however, for larger values
of $M^2$ the importance of the pole-continuum contribution grows and
eventually becomes larger than the pole part. This is an additional
reason to restrict the analysis to small values for the Borel parameters.

We have evaluated  the  sum rules for the coupling constant 
computed with all diagrams of Fig. 2 and we have found that they are very stable. 
In order to avoid repetition we do not show the corresponding curves of 
$g_{\Theta n K}$ as afunction of the Borel mass squared $M^2$.  
We give only the values of the coupling 
extracted at $M^2 = 1.5$ GeV$^2$ and  $M'^2 = 1$ GeV$^2$  in Table I.  
In what follows we shall present our results for the coupling constant 
$g_{\Theta n K}$ obtained with the color connected diagrams only.  
In Fig. 3 we show the coupling, given 
by the solution of the sum rule  I A (\ref{soma1}), as a function of the 
Borel mass squared $M^2$. 
Different lines show different values of the continuum threshold $\Delta_N$. As 
it can be seen, $g_{\Theta n K}$ is remarkably stable with respect to variations both in
$M^2$ and in $\Delta_N$. In Fig. 4 we show the coupling obtained with the sum
rule  II A  
(\ref{soma2}). We find again fairly stable results which are very weakly
dependent on the continuum threshold. 
Comparing the results obtained with  I A and  II A, we 
see that they are consistent with each other within the errors inherent to the 
method of QCD sum rules. In Fig. 5 we show the results of the 
sum rule  I B. 
In Fig. 6 we present  the result of the sum rule  II B. The meaning of  the
different lines is the same as in the previous figures. 
The results are similar to the cases before.

In Table \ref{table1} we present a summary of our results for $g_{\Theta n K}$ giving
emphasis to the difference between the results obtained with all diagrams and with 
only the color-connected ones. For the continuum thresholds we have 
employed $\Delta_N = \Delta_K = 0.5$ GeV. 
\begin{table}
\begin{center}
\begin{tabular}{|c|c|c|}  \hline
case & $|g_{\Theta nK}|$ (CC) &  $|g_{\Theta nK}|$ (all diagrams)  \\
\hline
\hline
I A & 0.71 & 2.59  \\
\hline
II A & 0.82 & 3.59   \\
\hline
I B & 0.84 & 3.24  \\
\hline
II B & 0.96  & 4.48  \\
\hline
\end{tabular}
\caption{\label{table1} $g_{\Theta nK}$ for various cases}
\end{center}
\end{table}

For our final value of $g_{\Theta nK}$ we take an average of the
sum rules  IA-IIB. It is interesting to observe that the influence of
the continuum threshold is relatively small, especially when compared to
the corresponding two-point functions.

Considering the uncertainties in the continuum thresholds, in the
coupling constants $\lambda_{K,N,\Theta}$ and in the quark condensate we
we get an uncertainty of about 50\%. Our final result then reads:
\begin{eqnarray}
  |g_{\Theta nK}| \mbox{(all diagrams)} &=& 3.48 \, \pm 1.8\,,\nn\\
  |g_{\Theta nK}| \mbox{(CC)}&=& 0.83 \, \pm 0.42\,.
\label{gfinal}
\end{eqnarray}
Including all diagrams, the prediction for $\Gamma_\Theta$ is
then 13 MeV (652 MeV) for a positive (negative) parity pentaquark.
In the CC case we get a width of 0.75 MeV (37 MeV) for a positive (negative) 
parity pentaquark.  
One should keep in mind that higher dimensional condensates and higher
order perturbative corrections could increase 
this prediction for the width. The parity of the pentaquark has not been 
experimentally determined. The measured value of the width is around 5-10 MeV 
both in the Kn channel (considered here) and in the Kp channel.

We see that it is difficult to obtain the measured decay width for a negative
parity pentaquark.


\section{Summary and conclusions}

We have presented a QCD sum rule study of the decay of the 
$\Theta^{+}$ pentaquark  using a  diquark-diquark-antiquark scheme
with one scalar and one pseudoscalar diquark. 
Based on an evaluation of the relevant three-point function, 
we have  computed the coupling  constant $g_{\Theta nK}$. 
In the operator product expansion we have included all diagrams
up to dimension 5. In this particular type of sum rule a complication
arises from the pole-continuum transitions which are not exponentially
suppressed after Borel transformation and must be explicitly included.
The analysis was made for two different pole-continuum parametrisations
and in two different evaluation schemes. The results are consistent with
each other. In addition, we have tested the ideas presented in
\cite{oga,ricanari} by including only diagrams with color exchange.
Our final results are given in eq. (\ref{gfinal}).

We find that for a positive parity pentaquark a width much smaller than 
10 MeV would indicate a pentaquark which contains no color-singlet
meson-baryon contribution. For a negative parity pentaquark, even
under the assumption that it is a genuine 5-quark state, we can
not explain the observed narrow width of the $\Theta$.
In \cite{Stech} the pentaquark was investigated in a nonrelativistic 
quark model for a diquark-diquark-antiquark configuration 
with two scalar diquarks in a relative P-wave.
It was found that a narrow width is difficult to achieve, unless
the pentaquark has an uncommon spatial structure. 
It seems that from the theory side a very small pentaquark width is
not impossible, but that the pentaquark has indeed to be in a 
special configuration to explain the observed decay width.

\smallskip

{\bf Acknowledgements:} It is a pleasure to thank R.D. Matheus for
instructive discussions. M.E is grateful to FAPESP for financial support
(contract number 2004/08960-4) and also acknowledges the hospitality extended 
to him during his stay in Brazil. This work has been supported by CNPq and  
FAPESP (Brazil). 
\vspace{0.5cm}




\begin{figure} \label{fig1}
\centerline{\psfig{figure=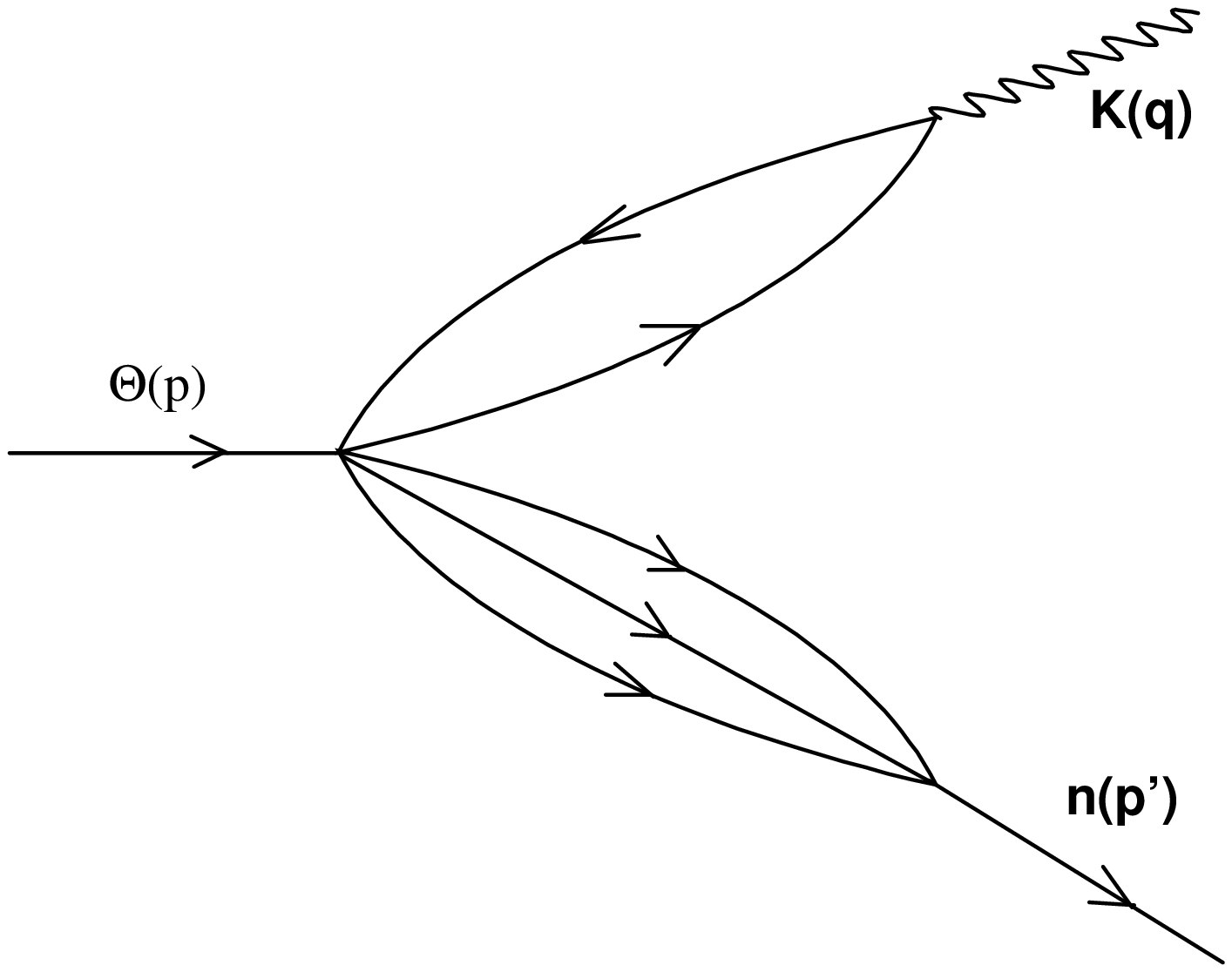,width=8cm,angle=0}}
\caption{The basic diagram for the theoretical side of the sum rule.}
\end{figure}

\begin{figure} \label{fig2}
\centerline{\psfig{figure=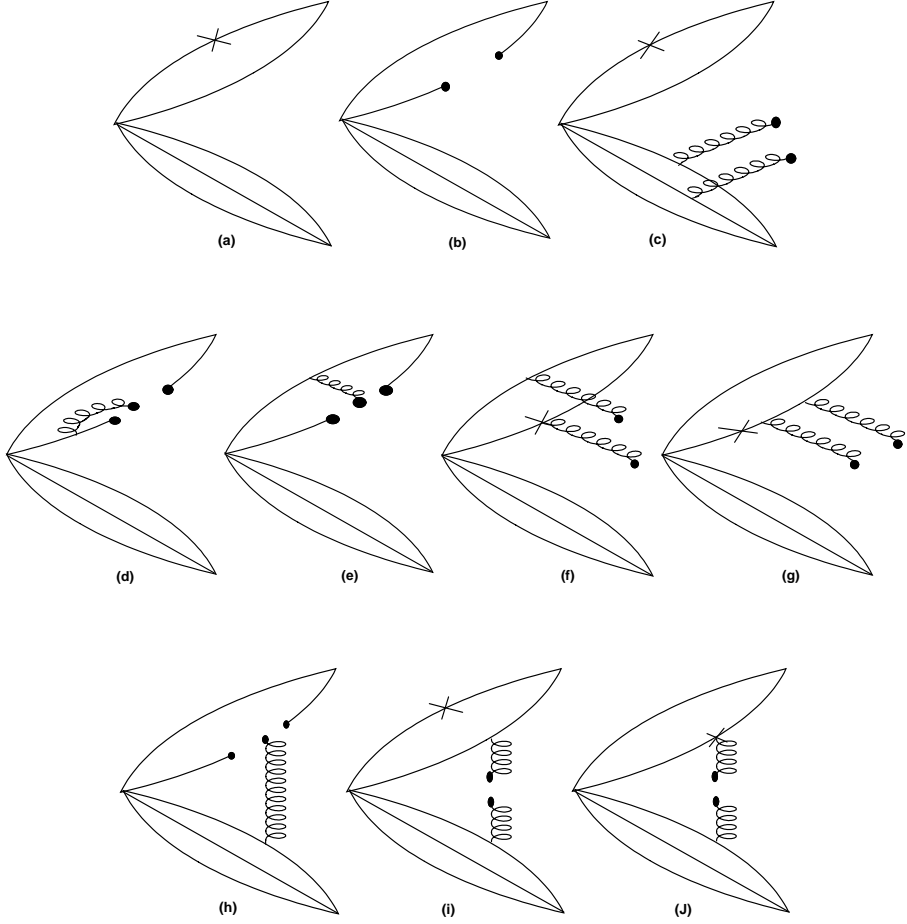,width=12cm,angle=0}}
\caption{The main diagrams which contribute to the theoretical side 
of the sum rule in the relevant structure.
a) - g) are the color disconnected diagrams, whereas h) - j) are the color 
connected diagrams. The cross indicates the insertion of
the strange mass.}
\end{figure}

\begin{figure} \label{fig3}
\centerline{\psfig{figure=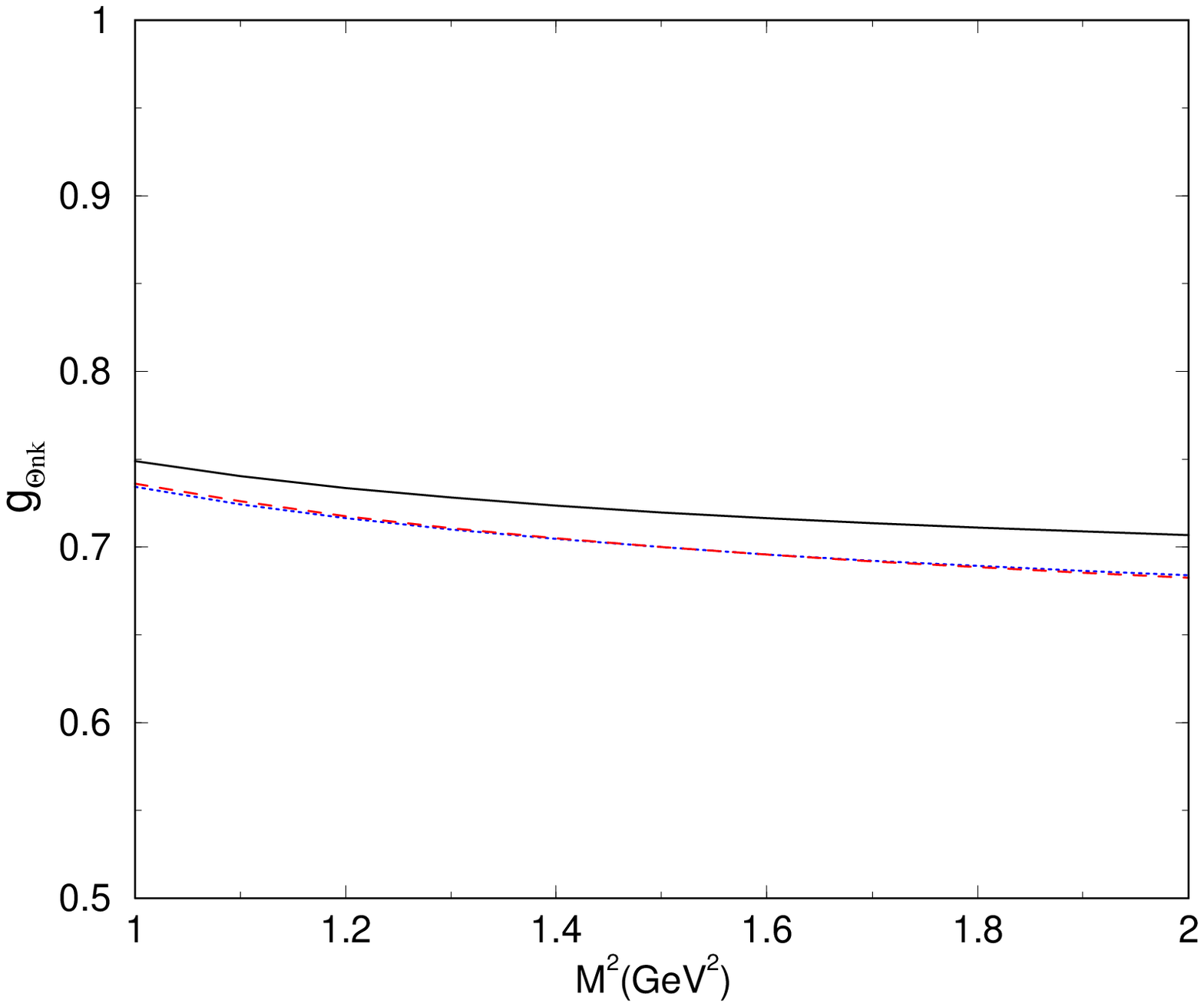,width=8cm,angle=0}}
\caption{$|g_{\tnk}|$ in case  I A with three different continuum threshold parameters. 
Solid line: $\Delta_N=0.5$ GeV, dotted line:$\Delta_N=0.4$ GeV, 
dash-dotted line:  $\Delta_N=0.6$ GeV.}
\end{figure}

\begin{figure} \label{fig4}
\centerline{\psfig{figure=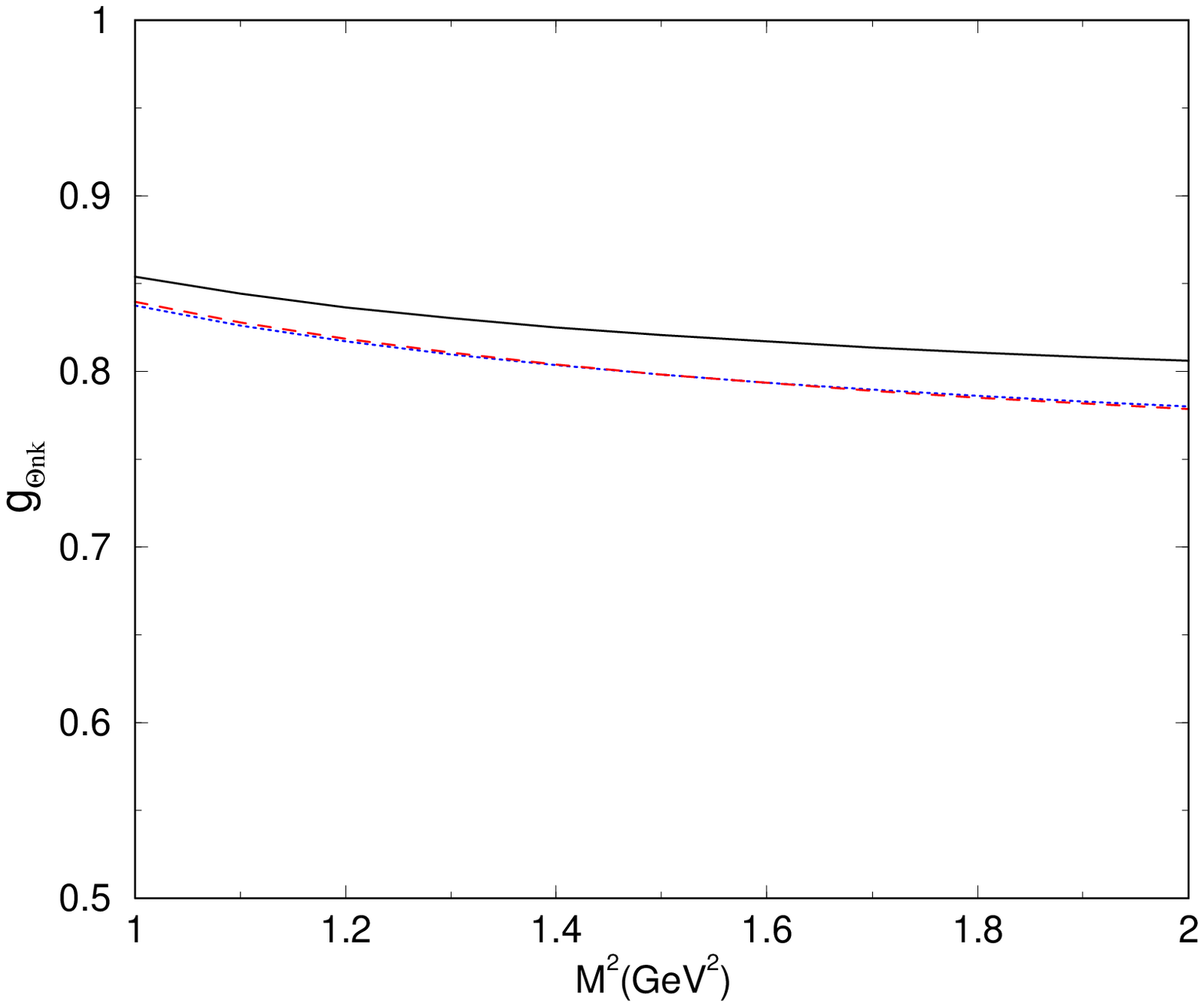,width=8cm,angle=0}}
\caption{$|g_{\tnk}|$ in case  II A. Solid line:  $\Delta_N=0.5$ GeV. 
Dotted line: $\Delta_N=0.4$ GeV. Dashed line: $\Delta_N=0.6$ GeV.
$M^{'2}=1$ GeV$^{2}$.} 
\end{figure}

\begin{figure} \label{fig5}
\centerline{\psfig{figure=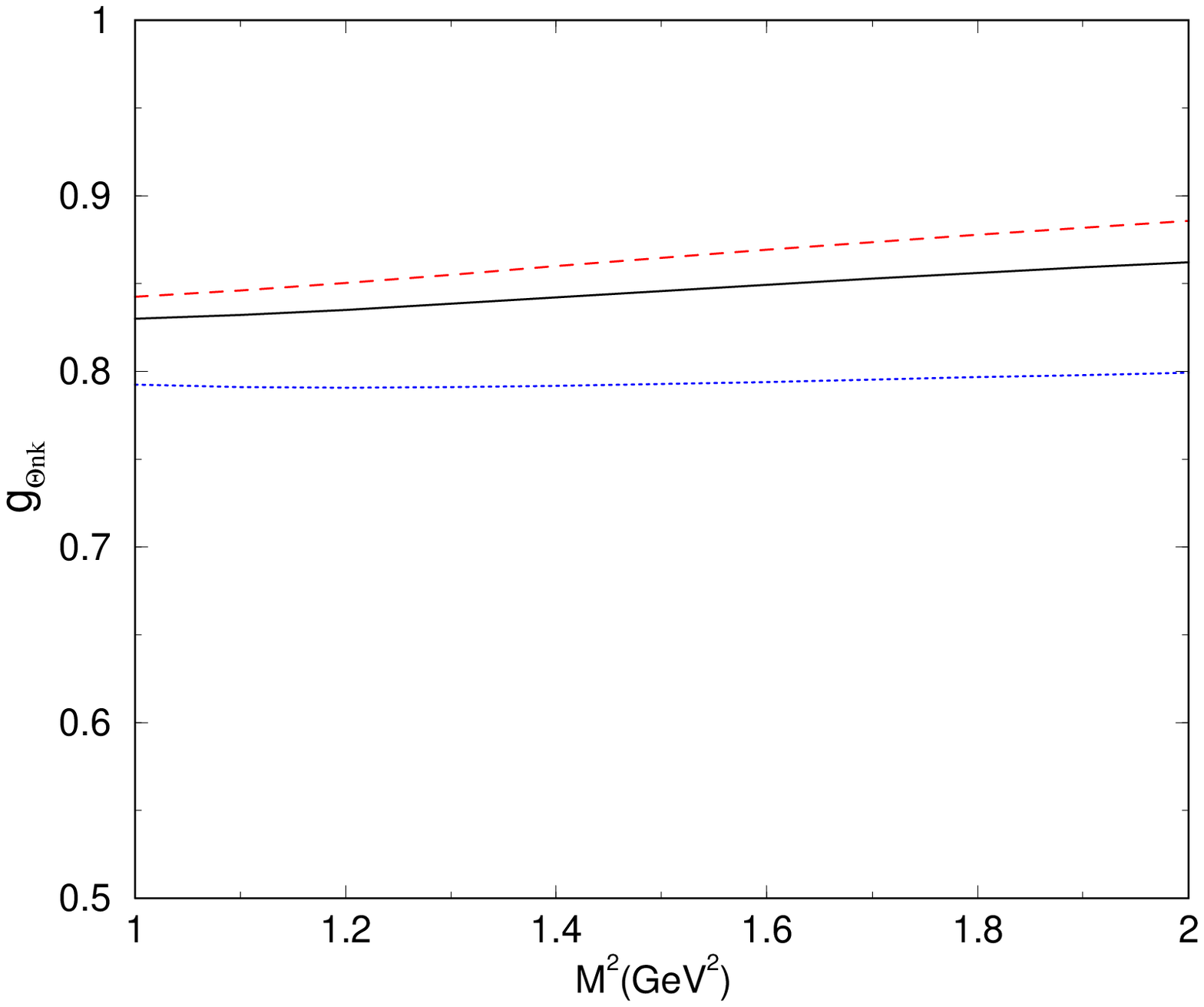,width=8cm,angle=0}}
\caption{$|g_{\tnk}|$ in case  I B  with three different continuum threshold parameters. 
Solid line: $\Delta_N=0.5$ GeV, dotted line:$\Delta_N=0.4$ GeV, 
dash-dotted line:  $\Delta_N=0.6$ GeV.}
\end{figure}

\begin{figure} \label{fig6}
\centerline{\psfig{figure=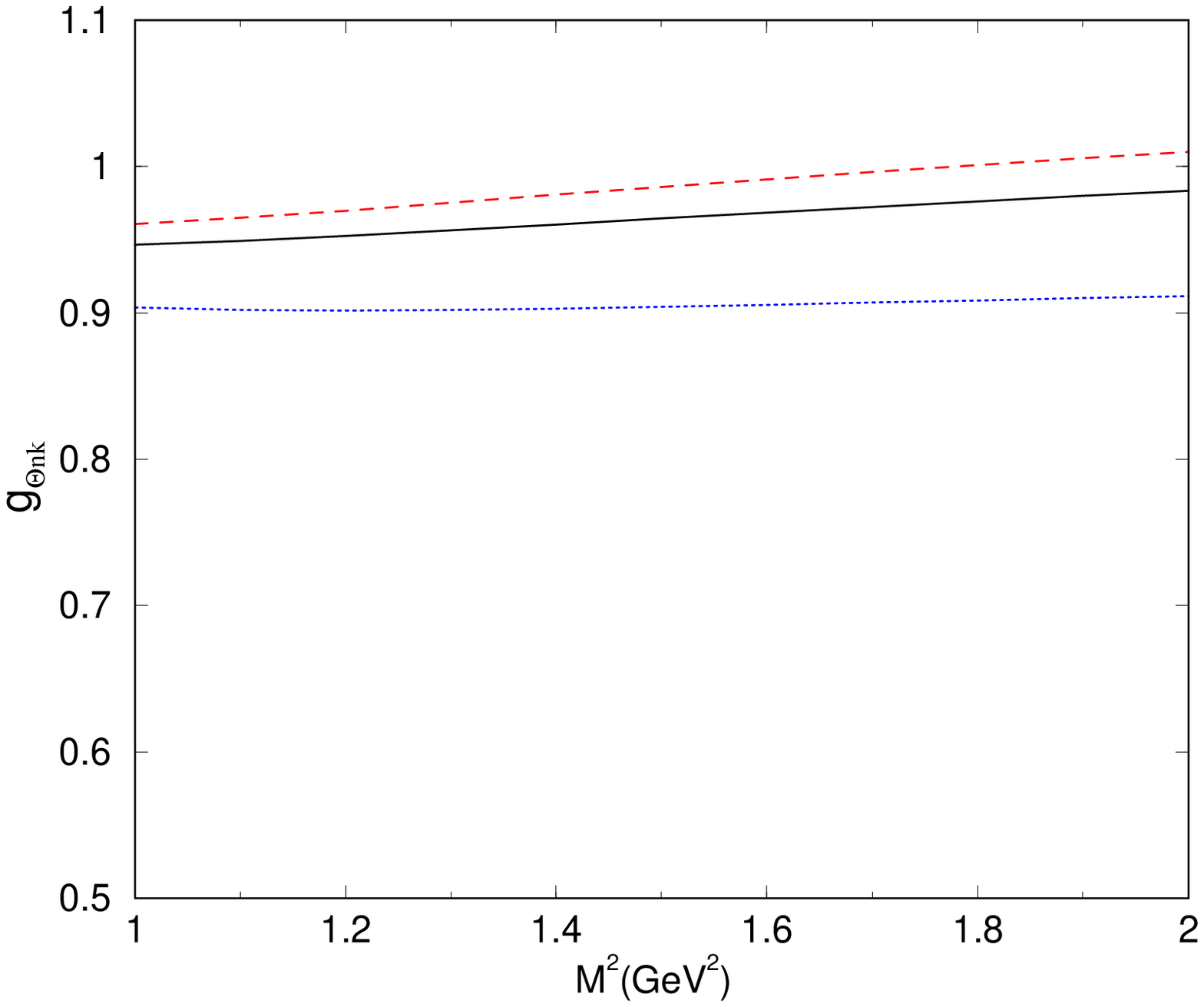,width=8cm,angle=0}}
\caption{$|g_{\tnk}|$ in case  II B. Solid line:  $\Delta_N=0.5$ GeV. 
Dotted line: $\Delta_N=0.4$ GeV. Dashed line: $\Delta_N=0.6$ GeV.
$M^{'2}=1$ GeV$^{2}$.} 
\end{figure}

\end{document}